\documentclass{aa}

\usepackage{graphicx}
\usepackage{txfonts}
\usepackage{braket}
\usepackage{amsmath}
\usepackage{bm}
\usepackage[colorlinks,allcolors=blue]{hyperref}
\usepackage[a]{esvect}
\makeatletter
\renewcommand*\aa@pageof{, page \thepage{} of \pageref*{LastPage}}

\makeatother
\usepackage{amstext}
\usepackage[normalem]{ulem}

\begin{document}

\title{Mitigating scanning-induced systematics in \textit{Hipparcos} and implications for \textit{Gaia} astrometry}

\author{
A. Binnenfeld\inst{1} \and
S. Shahaf\inst{2} \and
R.\ I.\ Anderson\inst{3} \and
H.-W. Rix\inst{2} \and
S. Zucker\inst{4,5}
}

\institute{
    Institute of Physics, Laboratory of Astrophysics, Ecole Polytechnique F\'{e}d\'{e}rale de Lausanne (EPFL), Observatoire de Sauverny, 1290 Versoix, Switzerland
    \and
    Max Planck Institute for Astronomy, Königstuhl 17, D-69117, Heidelberg, Germany
    \and
    Institut f\"ur Astrophysik und Geophysik, Georg-August-Universit\"at G\"ottingen, Friedrich-Hund-Platz 1, 37077 G\"ottingen, Germany
    \and
    School of Physics and Astronomy, Faculty of Exact Sciences, Tel Aviv University, Tel Aviv, Israel
    \and
    Department of Earth and Planetary Sciences, Faculty of Exact Sciences, Tel Aviv University, Tel Aviv 6997801, Israel}

\date{Accepted XXX. Received YYY}

\abstract{
Space-based astrometric missions such as \textit{Hipparcos} and \textit{Gaia} rely on one-dimensional scanning strategies that couple the measured signals with both observation times and scanning geometry. This coupling can introduce spurious periodicities and complicate their interpretation. We present a non-parametric extension of the phase-distance correlation (PDC) periodogram that quantifies and reduces dependence on scan geometry at the period-search stage. Using \textit{Hipparcos} data, we identified recurrent periodicities associated with characteristic scanning-law timescales and showed that they are strongly suppressed by the scan-angle-controlled statistic. We then applied the method to \textit{Gaia} Data Release 4 (DR4) prerelease epoch astrometry, where a recurrent $\sim$54-day feature is suppressed across multiple sources while independently known orbital signals remain detectable. The method provides a diagnostic of scan-angle dependence and a way to mitigate its impact on period searches. These results are particularly relevant for \textit{Gaia} DR4, where public epoch astrometry will enable large-scale searches for companions ranging from exoplanets to dormant black holes.
}


\titlerunning{Mitigating scanning-induced systematics in \textit{Hipparcos} and \textit{Gaia} astrometry}
\authorrunning{A. Binnenfeld et al.}

\maketitle
\section{Introduction}
\label{sec:intro}

Space astrometry missions have transformed our view of the Galaxy, providing precise positions, parallaxes, and proper motions for an enormous number of sources. \textit{Hipparcos} established the foundation of modern space astrometry \citep{1997A&A...323L..49P}, and \textit{Gaia} has extended it to more than a billion objects with much higher precision \citep{gaia2016, GaiaDR3summary}. Beyond its contribution to Galactic structure and stellar astrophysics studies, \textit{Gaia} is also a major tool for time-domain science \citep{eyer2023}, including the astrometric detection and characterization of stellar and substellar companions, as well as compact companions.

Astrometric measurements can broadly be divided into narrow-angle, or relative, astrometry and wide-angle, or absolute, astrometry. Narrow-angle astrometry measures the position of a target relative to nearby reference sources over a small field of view, whereas wide-angle astrometry aims to determine positions, parallaxes, and proper motions in a global reference frame over large angular scales \citep[see][]{LindegrenBastian}. Space missions such as \textit{Hipparcos} and \textit{Gaia} were designed for the latter purpose. Their scanning strategies allow measurements obtained at different times and in different parts of the sky to be linked into a global astrometric solution \citep{vanev98, gaia2016}.

A central feature of this observing mode is that these missions do not measure the full two-dimensional position of a source at each epoch. Instead, they repeatedly scan the sky following a complex scanning law, and each observation provides a one-dimensional positional measurement along the instantaneous scanning direction \citep{vanev98, gaia2016}. The astrometric time series of a given source therefore consists of pairs of along-scan (abscissa) measurements and scan orientations. This strategy is fundamental to the missions, but it also creates a non-trivial coupling between observation times, scan geometry, and the measured signals.

Given that the scan geometry is coupled to the temporal sampling, the periodogram of a scanning-astrometric time series reflects both the intrinsic variability of the source and the observing window imposed by the scanning law. This is analogous to the spectral-window problem in unevenly sampled time-series analysis \citep[see][]{VanderPlas18}, with the additional complication that each measurement is a one-dimensional projection along a scan direction that changes with time. As a result, scan-angle-dependent residuals or systematics can introduce apparent periodicities, and real periodic signals may be distorted or biased.

Recent work on \textit{Gaia} has shown that scan-angle signals can produce apparent orbital solutions and other false positives when not properly accounted for \citep{HOLL22, Halbwachs2023}. Spectroscopic follow-up has also revealed scan-law-related spurious periods in \textit{Gaia} Data Release 3 (DR3) astrometric candidates. For example, \citet{2023RNAAS...7...12S} find no evidence for the companion implied for HD 113283, whose period lies close to known \textit{Gaia} scanning-law artifacts. These effects are therefore a real limitation on the reliability of astrometric companion searches.

This problem is becoming increasingly important as \textit{Gaia} moves from a mission focused mainly on basic five-parameter astrometry to one that also delivers large catalogs of non-single-star (NSS) solutions. \textit{Gaia} DR3 already demonstrated the scientific potential of this transition by publishing an extensive NSS data set, including many solutions with orbital elements or acceleration parameters, and a very large sample of astrometric binary solutions \citep{GaiaDR3NSS, Halbwachs2023}, showing both the richness of the catalog and the need for careful treatment of selection effects and spurious solutions. 

\textit{Gaia} Data Release 4 (DR4) will include epoch astrometry, making it possible to detect and characterize astrometric companions. In particular, \textit{Gaia} is expected to become a major source of giant planets and substellar companions, with forecasts ranging up to tens of thousands of exoplanet detections \citep[e.g.][]{Lammers2026}. \textit{Gaia} is also proving powerful in the identification of unseen compact companions, including neutron stars and dormant black holes, catalogs expected to expand substantially with future releases \citep{Shahaf2019, Shahaf2024, ElBadry2024NS, Panuzzo2024}. Since these discoveries rely directly on weak orbital signatures in sparse astrometric time series, they are especially vulnerable to scanning-induced systematics.

In radial-velocity or photometric time-series analysis, the first step is often to search for periodicity without committing to a detailed physical model. Only after a candidate period is identified can a full orbital or variability solution be fitted. Periodograms are the standard tool for this purpose. Astrometric time series, however, are more complex than ordinary scalar measurements. As a result, orbit fitting is often performed directly, without a prior model-independent search for periodicity.

To address this gap, we introduced a model-independent periodogram for scanning astrometry \citep{Binnenfeld2023}, based on the phase-distance correlation approach (PDC; \citealt{zucker18, zucker19, 2024A&A...686A.192B}). The method enables the detection of periodic signals in one-dimensional astrometric data without assuming a specific orbital model.

Here, we extend the PDC framework to identify and suppress periodic signals associated with the observing geometry, building on our original periodogram framework. We primarily used \textit{Hipparcos} epoch measurements. Since \textit{Hipparcos} epoch astrometry is publicly available \citep{1997A&A...323L..49P, vanev98} and shares the essential one-dimensional scanning nature of \textit{Gaia} \citep{Lindegren2012, gaia2016}, it is highly useful for identifying, illustrating, and characterizing scanning-induced artifacts in real data. We then extend the discussion to \textit{Gaia}, where the same general problem is expected to play a central role in the interpretation of epoch astrometry and orbital solutions, especially in the context of DR4 companion searches.

The paper is organized as follows. In Sect.~\ref{sec:method}, we describe the method at the center of this work. In Sect.~\ref{sec:hip}, we demonstrate it using \textit{Hipparcos} data and identify common spurious periodic signals. In Sect.~\ref{sec:gaia}, we present a corresponding demonstration using public \textit{Gaia} data and discuss the implications for companion detection. Finally, in Sect.~\ref{sec:summary}, we summarize our results and discuss their relevance for future studies with \textit{Gaia}. We also briefly discuss additional applications for the framework.
\section{PDC for one-dimensional scanning astrometry}
\label{sec:method}

We presented the PDC periodogram for one-dimensional scanning astrometry in \citet{Binnenfeld2023}. In that work, each astrometric measurement, specified by its abscissa $v$ and scanning angle $\theta$, was represented geometrically in the plane, and a pairwise dissimilarity was defined to reflect the relative geometry of two such measurements. This allows us to construct a dissimilarity matrix for the astrometric data and apply the PDC framework without assuming a specific orbital model.

In \citet{Binnenfeld2023}, we referred to this dissimilarity as a metric. More precisely, the adopted expression does not in general satisfy the triangle inequality and is therefore not formally a metric. We retained the same dissimilarity here, noting that the distance-correlation framework can also be formulated in terms of dissimilarities \citep{sz_par_14}.

Distance correlation \citep{Szeetal2007} measures general statistical dependence and is sensitive to both linear and nonlinear relationships. This is particularly useful for scanning astrometry, where each observation is not a simple scalar measurement but also depends on its scanning direction. The PDC periodogram applies this idea by comparing the astrometric dissimilarity matrix with a phase-distance matrix, thereby quantifying the dependence between the measurements and the phase for each trial period \citep{zucker18}.

In this work, we extended that framework to account explicitly for scan-angle dependence. We introduced a separate distance between scanning directions, independent of the measured abscissae, and used it to construct a scan-angle-controlled statistic following the formalism of partial distance correlation \citep{sz_par_14, Binnenfeld2022}. In analogy with classical partial correlation \citep{Pea1915}, the contribution associated with scan angle is removed before evaluating the dependence between the astrometric measurements and phase.

Concretely, in addition to the astrometric dissimilarity matrix $A_{ij}$ and the phase distance matrix $B_{ij}$, as defined in \citet{Binnenfeld2023}, we define a third matrix $C_{ij}$ that encodes the distance between the scanning angles:
\begin{equation}
C_{ij} = d_{\mathrm{scan}}(\theta_i, \theta_j) \ .
\end{equation}
Since the scanning direction is treated as an unoriented line, we take the scan-angle coordinate to be periodic modulo $\pi$. We therefore define
\begin{equation}
\phi^{\theta}_{ij} = (\theta_i - \theta_j) \bmod \pi \ ,
\end{equation}
and use the PDC-style circular distance
\begin{equation}
d_{\mathrm{scan}}(\theta_i, \theta_j)
=
\phi^{\theta}_{ij}
\left(
\pi - \phi^{\theta}_{ij}
\right) \ .
\end{equation}

Let $\widetilde{A}$, $\widetilde{B}(P)$, and $\widetilde{C}$
denote the U-centered astrometric, phase, and scan-angle
dissimilarity matrices, respectively. For a generic matrix $D$, its
U-centered form is
\begin{equation}
\widetilde{D}_{ij}
=
D_{ij}
-\frac{D_{i\cdot}}{n-2}
-\frac{D_{\cdot j}}{n-2}
+\frac{D_{\cdot\cdot}}{(n-1)(n-2)},
\qquad i\neq j,
\end{equation}
with $\widetilde{D}_{ii}=0$, where
$D_{i\cdot}=\sum_k D_{ik}$,
$D_{\cdot j}=\sum_k D_{kj}$, and
$D_{\cdot\cdot}=\sum_{k,l}D_{kl}$. We use the U-centered inner
product
\begin{equation}
\left\langle \widetilde{D},\widetilde{E}\right\rangle_U
=
\frac{1}{n(n-3)}
\sum_{i\neq j}\widetilde{D}_{ij}\widetilde{E}_{ij},
\end{equation}
and the corresponding norm
$\|\widetilde{D}\|_U=
\sqrt{\langle\widetilde{D},\widetilde{D}\rangle_U}$.

To reduce the component associated with scan geometry, we project the
astrometric matrix onto the orthogonal complement of
$\widetilde{C}$,
\begin{equation}
\widetilde{A}_{\perp C}
=
\widetilde{A}
-
\frac{
\left\langle\widetilde{A},\widetilde{C}\right\rangle_U
}{
\left\langle\widetilde{C},\widetilde{C}\right\rangle_U
}
\widetilde{C},
\end{equation}
and define the scan-angle-controlled, partial PDC as
\begin{equation}
{\rm PDC}_{\rm partial}(P)
=
\frac{
\left\langle\widetilde{A}_{\perp C},
\widetilde{B}(P)\right\rangle_U
}{
\|\widetilde{A}_{\perp C}\|_U\,
\|\widetilde{B}(P)\|_U
}.
\end{equation}
Because only the astrometric matrix is residualized, this is formally a semi-partial rather than a full partial distance correlation. It measures the remaining phase dependence after this projection. For simplicity, we refer to this scan-angle-controlled semi-partial statistic as the partial PDC throughout the paper. The statistic is undefined if $n<4$, if $\|\widetilde{C}\|_U=0$, or if either norm in the final denominator vanishes.
This projection should not be interpreted as full statistical conditioning. In particular, suppression or survival of a periodic signal does not by itself fully establish whether its physical origin is instrumental or astrophysical.

This projection reduces the component of the periodogram associated with scan-angle dependence while increasing sensitivity to other periodic signals.
As in \citet{binnenfeld20}, the significance of detected periods can be assessed using either permutation tests or analytic approximations. In this work, we adopted the chi-square approximation proposed by \citet{fap_shen}, which enables an analytical estimation of false-alarm probabilities.

\section{\textit{Hipparcos}}
\label{sec:hip}

\subsection{\textit{Hipparcos} intermediate astrometric data (IAD)}

During its four-year mission, \textit{Hipparcos} measured positions, parallaxes, and proper motions for about $120\,000$ stars, reaching a precision of up to a few milliarcseconds (mas), unprecedented at the time. The resulting \textit{Hipparcos} catalog was published in 1997 \citep{1997A&A...323L..49P}. To test and demonstrate our method, we used the publicly available \textit{Hipparcos} Intermediate Astrometric Data (IAD). The data provide the individual abscissa measurements, specifically as the residuals after subtraction of the basic astrometric solution. Each observation is associated with a scanning angle and time, forming a sparse and unevenly sampled one-dimensional astrometric time series.

Two independent reductions of the IAD are available, produced by the FAST \citep{FAST} and NDAC \citep{NDAC} consortia. While both are based on the same raw observations, they differ in details of the data processing and calibration. 
We used the FAST reduction for the analysis presented below. As a robustness check, we repeated the calculation using NDAC and compared the resulting periodograms, including the recurrence and suppression of the characteristic features discussed in Sect.~\ref{spur}. The population-level results were consistent between the two reductions, although individual source periodograms were not fully identical.

In applying our method, we also specified the parameter $L$, which represents the effective length of the segment on which the true astrometric position is assumed to lie. We tested $L=0.01,0.03,0.1,0.3,1,$ and $3$~mas using a random sample of 5\,000 sources. Over that range, the median periodogram correlations exceed 0.9998, at least 97.2\% of the dominant peaks are in agreement, and the recurrent feature counts change by less than 1\%. We therefore adopted $L=1$~mas as a representative value within this stable range. This value specifically matches the \textit{Hipparcos} sample, for which $L=1$ mas is comparable to the typical scale of the along-scan residuals.

We define $\mathcal{D}_{\mathrm{cpl}}$ as the normalized inner product between the $\mathcal{U}$-centered astrometric and scan-angle distance matrices,
\begin{equation}
\mathcal{D}_{\mathrm{cpl}}
=
\frac{
\left\langle \widetilde{A}, \widetilde{C} \right\rangle_U
}{
\left\|\widetilde{A}\right\|_U
\left\|\widetilde{C}\right\|_U
}.
\end{equation}

\begin{figure}
\includegraphics[width=1\columnwidth,clip=true]{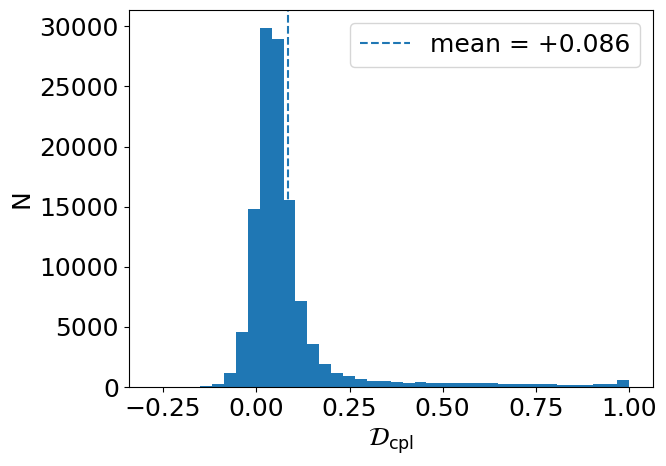}
\caption{Distribution of $\mathcal{D}_{\mathrm{cpl}}$ for all \textit{Hipparcos} sources. The catalog has a mean of 0.09 and standard deviation of 0.15, with an evident tail indicating a subset of sources showing strong dependence on scan angle.}
\label{fig:ast_part_dist}
\end{figure}

In practice, $\mathcal{D}_{\mathrm{cpl}}$ is useful as a fast diagnostic: it avoids scanning over trial periods, as in a periodogram, and does not require fitting a full orbital model. It therefore allows us to quickly identify sources whose astrometric residuals show a strong dependence on scan angle, for which scan-geometry effects are likely to be relevant in the subsequent period search. We computed $\mathcal{D}_{\mathrm{cpl}}$ for all sources in the sample, as shown in Fig. \ref{fig:ast_part_dist}. The catalog had a mean $\mathcal{D}_{\mathrm{cpl}}$ of $0.086$ and a standard deviation of $0.155$, with a pronounced high-value tail. For the exploratory analysis below, we focused on the $4\,460$ sources with $\mathcal{D}_{\mathrm{cpl}} > 0.5$. This threshold was not used as a formal detection criterion, but rather as a practical way to select sources in the high-dependence tail, where scan-angle effects are expected to be most relevant.

We note that $\mathcal{D}_{\mathrm{cpl}}$ is based on a normalized inner product between centered distance matrices, and can therefore take both positive and negative values. Positive values indicate stronger coupling between the astrometric residuals and the scan geometry, while negative values do not necessarily carry an opposite meaning \citep[see][]{sz_par_14}.

\subsection{Common spurious signals}
\label{spur}

Applying the standard PDC periodogram to the above sample reveals a set of prominent recurring signals at periods of approximately $\mathord\sim 28$, $\mathord\sim 49$, and $\mathord\sim 82$ days. These signals appear repeatedly among sources with strong scan-angle dependence, with the $\mathord\sim 49$-day peak being the most prominent in our inspection, while the $\mathord\sim 28$ and $\mathord\sim 82$-day signals are also quite frequently present.

To quantify this, we searched all 4\,460 sources in the $\mathcal{D}_{\mathrm{cpl}}>0.5$ sample. A feature was counted when a periodogram peak occurred within 5\% of its characteristic period and had a pointwise significance of $p<0.01$. The features centered at 28.50, 49.31, and 82.86~d were detected in 1\,717 (38.5\%), 3\,513 (78.8\%), and 1\,816 (40.7\%) sources, respectively. After accounting for scan angle using our method, 1\,048 (61.0\%), 3\,309 (94.2\%), and 1\,545 (85.1\%) of the corresponding detections no longer met the same significance criterion. As a control, we applied the same criterion to a reproducible random sample of 5\,000 sources with $\mathcal{D}_{\mathrm{cpl}}\leq0.5$, finding only 35 (0.70\%), 18 (0.36\%), and 16 (0.32\%) cases, respectively. Thus, these features are not entirely absent below the coupling threshold, but are strongly concentrated among sources with high scan-angle coupling.

The recurrence of these signals across the sample strongly suggests that they are not astrophysical in origin, but instead reflect effects associated with the observing strategy. These characteristic timescales are naturally connected to the $\mathord\sim57$-day precession period of the \textit{Hipparcos} scanning law and to the annual modulation of the observing geometry. In frequency space, the scanning window can generate aliases at combinations of the form $n f_{\rm prec} \pm m f_{\rm yr}$, where $f_{\rm prec}=1/P_{\rm prec}$ and $f_{\rm yr}=1/{\rm yr}$. The $\mathord\sim28$-day signal is consistent with the first harmonic of the precession period, $P_{\rm prec}/2$, while the $\mathord\sim49$-day signal is close to the annual alias $(f_{\rm prec}+f_{\rm yr})^{-1}$. The $\mathord\sim82$-day signal can also be associated with a low-order alias, approximately $(f_{\rm prec}-2f_{\rm yr})^{-1}$. We therefore interpret these recurrent periods as scanning-induced features arising from the coupling between scan geometry and temporal sampling, rather than as astrophysical periods.

As a robustness check, we repeated this analysis using the NDAC reduction for the above sources. The FAST and NDAC classifications of the approximately 28, 49, and 82~d features agreed for 89.6\%, 95.9\%, and 93.4\% of the sources, respectively, and the corresponding suppression fractions differed by no more than 2.9 percentage points. Thus, the population-level results are robust to the choice of reduction.

\begin{figure}
\includegraphics[width=1\columnwidth,clip=true]{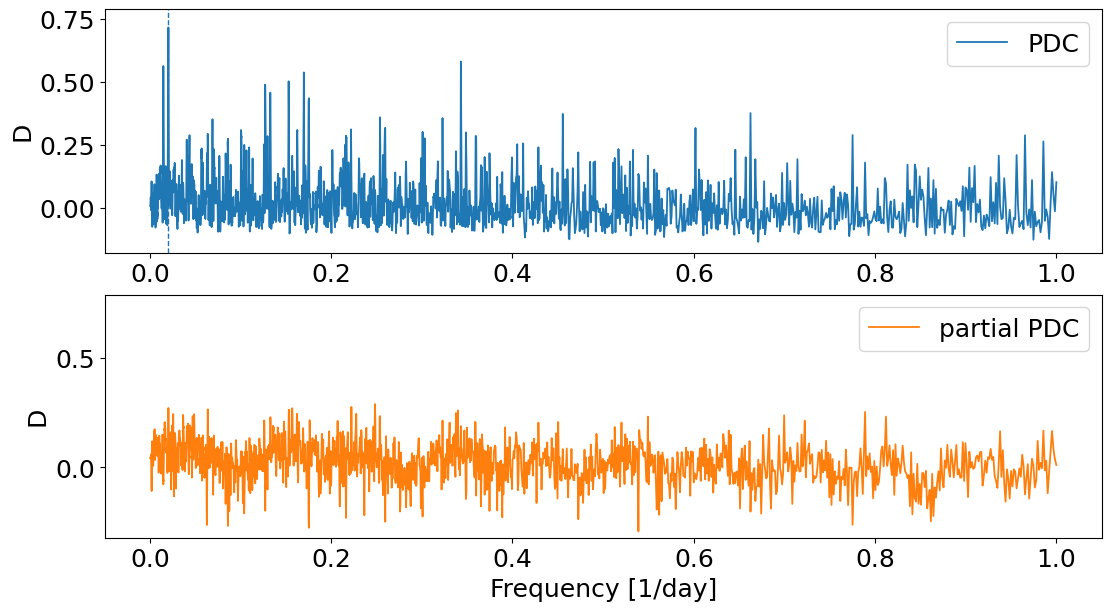}
\caption{Standard (top) and partial (bottom) PDC periodograms for HIP 29616. A prominent peak at $f=0.0204\,{\rm d}^{-1}$, corresponding to $P\simeq49.1$ days, is clearly visible in the standard PDC, consistent with a repeating signal driven by scan-angle systematics. This feature is effectively suppressed in the partial PDC.}
\label{fig:ast_part_example1}
\end{figure}

\begin{figure}
\centering
\includegraphics[width=1\columnwidth,clip=true]{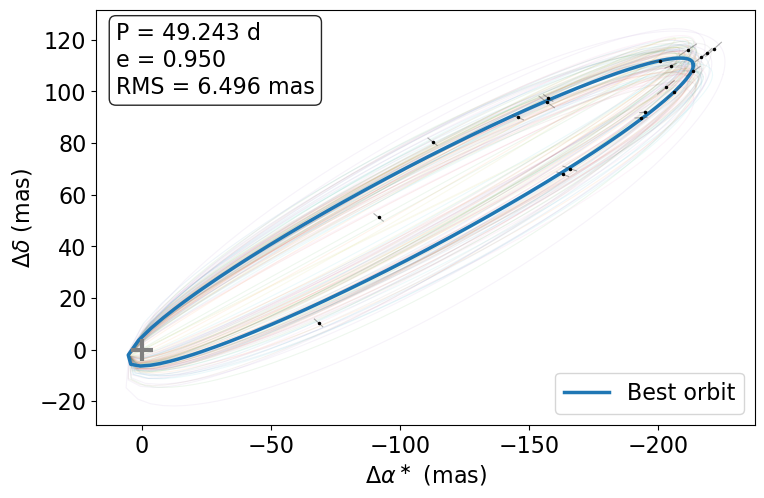}
\caption{
Best-fit astrometric orbit for HIP 29616 at the spurious $P\simeq49$ d period. The blue curve shows the best-fitting solution, and the along-scan measurements are overplotted. The faint curves show 80 realizations obtained by varying the fitted linear orbit coefficients within their estimated uncertainties, while keeping $P$, $e$, and $\tau$ fixed. The best-fitting solution has $P=49.243$ d, reaches the imposed eccentricity limit at $e=0.950$, and has an RMS residual of $6.496$ mas.
}\label{fig:data29616_1}
\end{figure}

\begin{figure}
\centering
\includegraphics[width=1\columnwidth,clip=true]{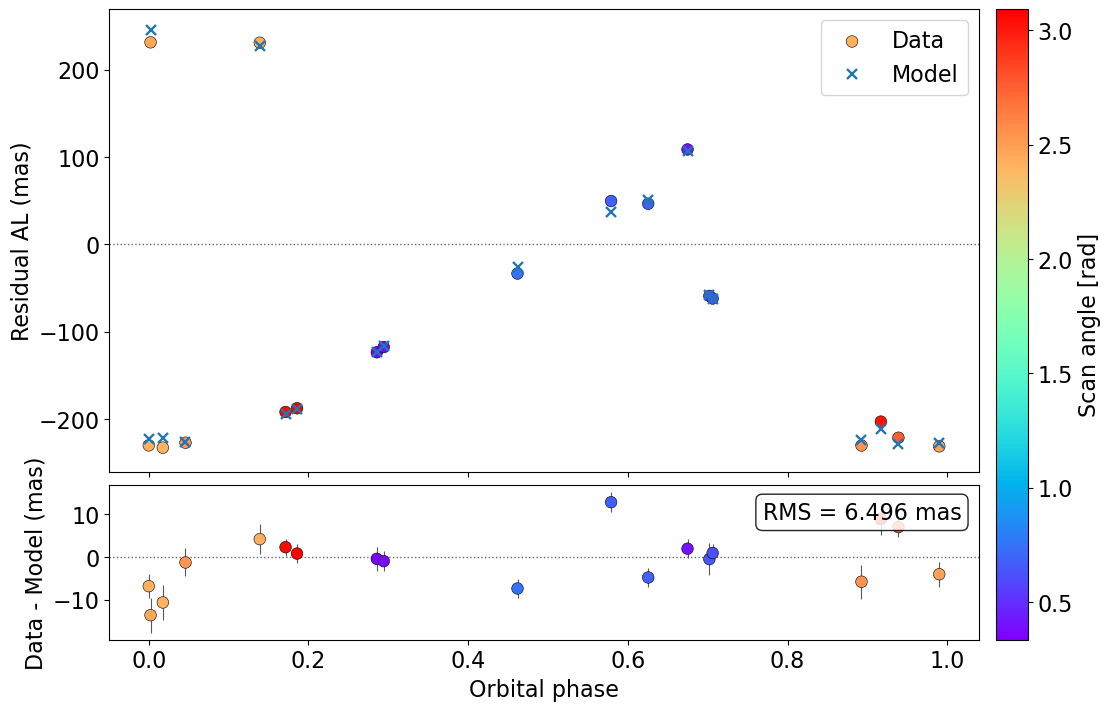}
\caption{Along-scan measurements of HIP 29616 phase-folded at $P\simeq49$ d. Top: along-scan residuals with the best-fit astrometric model overlaid. Points are colored by scan angle. Bottom: data-model residuals, with an RMS of $6.496$ mas.}
\label{fig:data29616_2}
\end{figure}

A representative example is shown in Fig.~\ref{fig:ast_part_example1},
where the standard and partial periodograms are compared for HIP~29616
(HD~42954). The spurious peak at the characteristic timescale is clearly visible in the standard PDC and disappears in the partial formulation. The underlying data, along with a Keplerian fit at this period, are shown in Fig.~\ref{fig:data29616_1} and Fig.~\ref{fig:data29616_2}. 

We fitted a Keplerian astrometric model to the 19 FAST abscissa residuals of HIP~29616, adopting the reported measurement uncertainties. The period $P$, eccentricity $e$, and periastron phase $\tau$ were treated as nonlinear free parameters, while the four Thiele--Innes coefficients and a constant along-scan offset were solved by weighted linear least squares at each trial. The best-fit solution was defined as the minimum-$\chi^2$ solution and yielded $P=49.243$~d and $e=0.95$, with the eccentricity reaching the imposed upper bound.

Taken at face value, this fit would imply an extremely eccentric, short-period orbit. The pericenter separation would be $q=a(1-e)$, corresponding to only a few stellar radii for a typical stellar-mass binary. Such an orbit would imply an unusually small pericenter separation for a stellar binary. 
Together, the recurrence of this period across the sample, its connection to the scanning-law timescales, its strong suppression after accounting for scan angle, and the unusual orbital interpretation for HIP~29616 support the conclusion that this particular signal is dominated by scan-geometry effects. This example illustrates how a scan-angle-associated periodicity can lead to an apparently well-defined but astrophysically misleading Keplerian fit, and how the proposed statistic can flag and mitigate such signals at the period-search stage.

\section{Application to \textit{Gaia} data}
\label{sec:gaia}

The June 2026 \textit{Gaia} DR4 astrometry prerelease provides epoch measurements for 12 selected \textit{Gaia} sources.\footnote{\url{https://www.cosmos.esa.int/web/gaia/dr4-prerelease}}
For each source, we fitted and subtracted a standard five-parameter astrometric model and computed the ordinary and scan-angle-controlled PDC periodograms over trial periods from 1 to 3000~d.

Since the \textit{Gaia} residuals are substantially smaller than those of \textit{Hipparcos}, we tested $L=0.001$--$1$ mas separately for the prerelease sample. Values up to $L\leq0.03$ mas yield very similar periodograms. We therefore adopted $L=0.01$ mas as a representative value within this range, consistent with the smaller characteristic scale of the \textit{Gaia} along-scan residuals.

\begin{figure*}
    \centering
    \includegraphics[width=\textwidth]{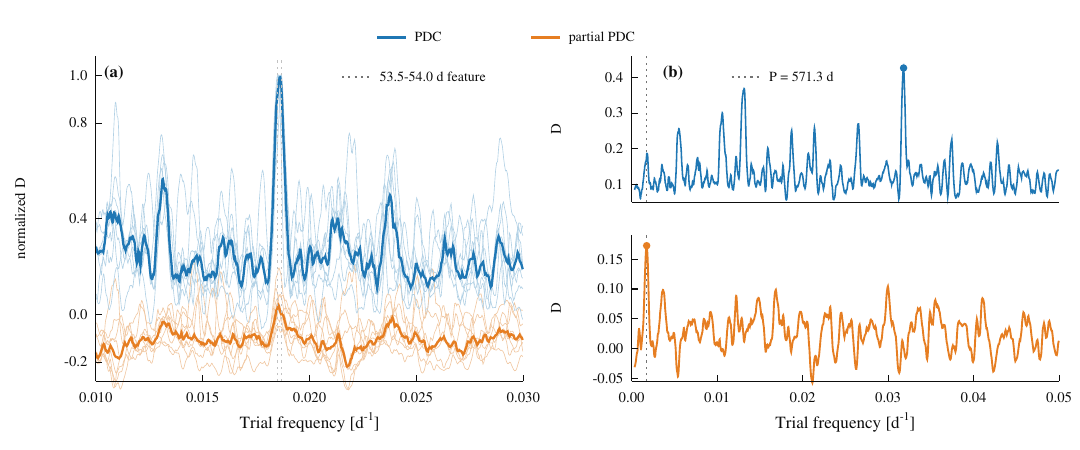}
    \caption{
    Application to the \textit{Gaia} prerelease epoch astrometry.
    Left: PDC (blue) and partial PDC (orange) periodograms around the recurrent scanning-law feature. Thin curves show the eight affected comparison sources, normalized by their PDC peak values, while thick curves show their median. The dotted lines mark the $53.5$--$54.0$ d feature, which is strongly suppressed by the partial PDC.
    Right: PDC (top) and partial PDC (bottom) periodograms of \textit{Gaia}-4. While the PDC is dominated by a higher-frequency peak, the partial PDC recovers the known orbital frequency corresponding to $P=571.3$ d (vertical dotted line).
    }
    \label{fig:gaia_prerelease}
\end{figure*}

Among the nine prerelease sources not presented in the release as orbital examples, eight show their strongest ordinary-PDC peak within the narrow range 53.51--54.03~d (Fig.~\ref{fig:gaia_prerelease}). This closely matches the $53.7\pm1.1$~d scan-angle-related feature identified in \citet{HOLL22}, supporting the interpretation of the recurrent peak as a scanning-law artifact. Accounting for scan angle strongly suppresses the feature in all eight sources.

The prerelease also contains systems with independently known orbital periods. For \textit{Gaia}-4, the ordinary PDC is dominated by a 31.44-d peak, whereas the scan-angle-controlled periodogram peaks at 569.93~d, close to the known 571.3-d orbit \citep{Stefansson2025}. For HD~114762, both formulations recover the known ${\sim}84$-d period \citep{Latham1989}. These examples illustrate that known astrophysical signals can remain detectable after removing the scan-angle component. We note that since the prerelease sample is small and curated, these results should be viewed as an empirical demonstration, and they do not necessarily reflect the frequency of such effects in the full \textit{Gaia} population.

We also applied the method to the epoch astrometry of \textit{Gaia} BH3 \citep{Panuzzo2024}, which is one of the few examples of \textit{Gaia} epoch astrometry that was publicly released before \textit{Gaia} DR4. The standard PDC shows a clear low-frequency maximum associated with the long-timescale astrometric motion of the system, and accounting for the scan angle leaves this signal essentially unchanged. This therefore provides an example of a valid astrophysical signal that is preserved by the partial formulation. Since the available ${\sim}5$-yr baseline covers only part of the orbit, the low-frequency maximum traces the long-term astrometric variation rather than the true orbital period, as expected for PDC when the period exceeds the observing baseline \citep{Barbey25}.

\section{Conclusion and future outlook}
\label{sec:summary}

\subsection{Main conclusions}

We extended the PDC periodogram to explicitly account for scan-angle dependence in one-dimensional astrometry.
By incorporating the scanning direction through partial distance correlation, the method suppresses signals associated with the observing geometry while retaining sensitivity to periodic variability not dominated by scan-angle effects. Because it operates before any orbital model is fitted, it provides a non-parametric, scan-aware diagnostic at the period-search stage, complementing subsequent physical modeling.

Applied to Hipparcos, the method suppresses recurrent periodicities associated with the scanning law while preserving signals not dominated by scan-angle dependence. The same is relevant for \textit{Gaia} epoch astrometry, where the sparsely sampled one-dimensional measurements make scan-aware period detection particularly useful for large-scale companion searches.

The scan-angle-controlled PDC should be interpreted as a diagnostic of sensitivity to the observing geometry rather than as a classifier of physical origin. Genuine astrophysical signals may be suppressed when they are coupled to scan geometry through projection and temporal sampling, while instrumental signals that are not strongly scan-angle dependent may remain. More generally, the projection used here is not equivalent to full statistical conditioning \citep[see][]{sz_par_14}, so establishing the physical origin of a candidate signal still requires, where appropriate, explicit physical modeling.

\subsection{Additional applications}

The same framework could also be applied to nuisance variables other than scan angle. One natural example is epoch photometry, which could be useful for variability-induced movers (VIMs), where measured astrometric motion is caused by changes in the brightness of unresolved components rather than by true orbital motion \citep{1996A&A...314..679W}. In such cases, accounting for photometric variability could test whether the astrometric signal is associated with luminosity-driven photocenter shifts. In \textit{Gaia} DR4, the availability of epoch astrometry, radial velocities, and multi-band photometry will make this kind of joint analysis possible.

Intrinsically variable giant stars provide a related application. In such objects, the astrometric signal does not necessarily originate from orbital motion or scanning-law systematics alone. Evolved stars with radii of order au are not strictly point-like for astrometric purposes: even when they are not spatially resolved, surface-brightness asymmetries across their stellar disks can displace the photocenter by a non-negligible fraction of the stellar radius, producing astrometric perturbations relevant for high-precision \textit{Gaia} measurements \citep{2011A&A...532A..13P}. This effect is expected to be especially relevant for evolved stars with large angular diameters, such as Mira variables, Cepheids, asymptotic-giant-branch stars, and red supergiants \citep{2011A&A...528A.120C}. In these cases, large-scale surface-brightness asymmetries, pulsation, shocks, and convective structures can shift the measured photocenter of the star relative to its barycenter.

Scanning astrometry is naturally sensitive to such effects because each observation measures a one-dimensional projection of the instantaneous photocenter along a different scan direction. If the photocenter motion is real, rather than induced by the scanning law, observations obtained at different position angles should contain mutually consistent information about the two-dimensional motion of the photocenter on the sky. Conversely, if the apparent signal is dominated by scan-angle-dependent systematics, its dependence on phase should be strongly reduced when controlling for the scanning geometry. The partial-PDC framework therefore provides a way to assess whether apparent photocenter variability is coupled to the scanning geometry.

Other effects can also produce apparent astrometric variability without requiring orbital motion of the measured source. One example is chromatic variability. In long-period variables (LPVs) and other large-amplitude variable stars, periodic color changes can modify the effective chromatic response of the instrument, and therefore the shape and centroid of the line-spread function, even for unresolved sources \citep{2006A&A...449..827B, 2021A&A...648A..44M}. If the photometric or chromatic variability period is known, for example from \textit{Gaia} epoch photometry, it could be included as an additional variable in the partial-PDC framework, either for individual sources or for ensembles of stars with similar variability properties.

Another class of cases involves blending and source structure. In visual binaries or crowded fields, the measured along-scan position can depend on the scan angle through the projected separation and flux contrast of neighboring sources \citep{HOLL22}. Such effects can also interact with photometric variability when changes in the flux ratio of unresolved components shift the photocenter preferentially along a particular direction \citep{1996A&A...314..679W, Halbwachs2023}. More generally, the same partial-PDC framework can be extended beyond scan-angle systematics by testing whether an apparent astrometric periodicity remains after accounting for the relevant observational, photometric, or geometric factors.

\section*{Data availability}

The Hipparcos and Gaia data analyzed in this work
are publicly available through the respective
mission archives. The Python implementation
of the periodograms is publicly available
as part of the SPARTA package \citep{SPARTA2020},
at \url{https://github.com/SPARTA-dev/SPARTA}.

\begin{acknowledgements}
We thank the anonymous referee for a careful and constructive report that substantially improved the manuscript.
This work has made use of data from the European Space Agency (ESA) mission \textit{Gaia} (\url{https://www.cosmos.esa.int/gaia}), processed by the \textit{Gaia} Data Processing and Analysis Consortium (DPAC, \url{https://www.cosmos.esa.int/web/gaia/dpac/consortium}). Funding for the DPAC has been provided by national institutions, in particular the institutions participating in the \textit{Gaia} Multilateral Agreement.
This research was supported by the Ministry of Innovation, Science \& Technology, Israel (grant 3-18143). S. Shahaf and H.-W. Rix acknowledge support from the European Research Council for the ERC Advanced Grant [101054731]. Funded by zukunft.niedersachsen, the joint science funding program of the Lower Saxony Ministry of Science and Culture and the Volkswagen Foundation.
The analyses in this paper made use of the following software packages: NumPy \citep{numpy}, SciPy \citep{2020SciPy}, and SPARTA \citep{SPARTA2020}.
\end{acknowledgements}

\bibliographystyle{aa}
\bibliography{pPDC}

\end{document}